\documentclass[a4paper,10pt]{article}
\usepackage{graphicx}
\title{Alteration of Chemical Concentrations through Discreteness-Induced Transitions in Small Autocatalytic Systems}
\author{Yuichi \textsc{Togashi}\footnote{E-mail: togashi@complex.c.u-tokyo.ac.jp}
 and Kunihiko \textsc{Kaneko}\\
{\footnotesize Department of Basic Science, School of Arts and Sciences, University of Tokyo, Komaba, Meguro, Tokyo 153-8902 Japan}}
\date{October 22, 2002}

\begin{document}
\sloppy
\maketitle
\begin{abstract}
We study an autocatalytic system consisting of several
interacting chemical species.
We observe a strong dependence of the concentrations of
the chemicals on the size of the system.
This dependence is caused by the discrete nature
of the molecular concentrations.
Two basic mechanisms responsible for them are identified and elucidated.
The relevance of the transitions
to processes in biochemical systems and in micro-reactors is
briefly discussed.
\end{abstract}
\begin{center}
\end{center}

Rate equations are often employed in the study of biochemical
reaction processes.
In rate equations, the quantities of chemicals are treated as continuous
variables,
and the actual
discreteness of the molecular concentration
is ignored.
Of course, fluctuations of numbers of molecules
have been studied using stochastic differential equations
and introduced non-trivial effects \cite{NIP,SR}.
Still, discreteness has not been considered in any such study.
In many biochemical processes, however,
some chemicals play important roles at extremely low concentrations,
amounting to only a few molecules per cell \cite{biology1,biology2}.
Furthermore, there exist amplification mechanisms involving enzymes
in cells through which
even a change by one molecule in a cell can result in
drastic effects.
In such situations,
the discreteness of the molecular concentration is
obviously not negligible.

We previously showed the existence of a novel transition
induced by the discreteness of the molecular concentration in
an autocatalytic reaction system \cite{YTKK}.
The system contains four chemicals $X_{i} (i=1,\cdots,4)$.
We considered an autocatalytic reaction network (loop)
represented by
$X_{i} + X_{i+1} \rightarrow 2X_{i+1}$ (with $X_{5}\equiv X_{1}$)
within a container
that is in contact with a reservoir of molecules.
Through interaction with the reservoir,
each molecule species $X_{i}$ diffuses in and out
at a total rate of $D_{i}s_{i}V$,
where $D_{i}$ is the flow rate,
$s_{i}$ the concentration of chemical $X_{i}$ in the reservoir,
and $V$ the volume of the container.
In this system, a novel state appears as a result of fluctuations and
the discreteness of the molecular concentration,
characterized as extinction and subsequent reemergence
of molecule species
alternately in the autocatalytic reaction loop.

\begin{figure}
\begin{center}
\includegraphics[width=70mm]{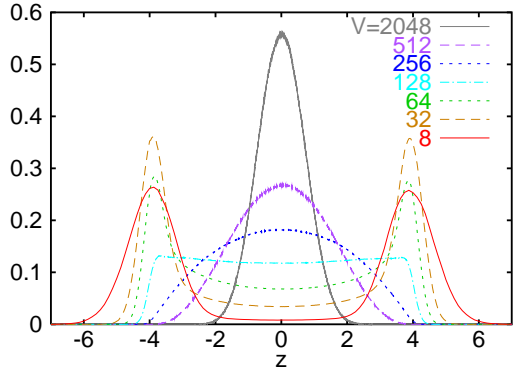}
\end{center}
\caption{Probability distribution of
$z \equiv (x_{1}+x_{3})-(x_{2}+x_{4})$.
Here $s_{i}=1$, and $D=1/128$.
For $V \ge 256$, $z$ has a distribution around $0$, corresponding
to the fixed point state $x_{i}=1$ (for all $i$).
For $V \le 64$, the distribution has double peaks around
$z=4$, corresponding to the 1-3 rich state
($N_{1}, N_{3} \gg N_{2}, N_{4} \approx 0$),
and $z=-4$, corresponding to the 2-4 rich state.}
\label{fig:dist-ac-bd}
\end{figure}

When the volume of the container is small,
$N_{i}$, the number of molecules of species $X_{i}$,
may go to $0$ (i.e. become extinct)
through a finite-size fluctuation due to the discreteness of the
molecular concentration.
Once $N_{i}$ reaches $0$, it remains $0$ until an $X_{i}$ molecule flows in.
Thus, if the flow rate of molecules is sufficiently small,
state with $N_{1}=N_{3}=0$ or $N_{2}=N_{4}=0$ can be realized.
In a state
with $N_2, N_4 \approx 0$ (a ``1-3 rich'' state),
switches between states with $N_1 > N_3$ and $N_3 > N_1$ can occur,
and similarly for a state with $N_1, N_3 \approx 0$ (a ``2-4 rich'' state).
A symmetry-breaking transition to these states was observed
in our previous study, with the decrease of $V$, as is shown in
Fig. \ref{fig:dist-ac-bd}, as the change of the probability distribution of
the number of molecules.
For large $V$, corresponding to the continuum limit,
the distribution of $z=((N_1+N_3)-(N_2+N_4))$ shows a single-peaked distribution
around $z=0$, whereas it is replaced by a symmetric, double-peak distribution as
$V$ is decreased.  This is a novel
discreteness-induced transition (DIT) occurring with the decrease
of $V$.  The transition occurs without any change of parameters, and thus
cannot be discussed in the rate equation with noise (i.e., by the continuum description).

\begin{figure}
\begin{center}
\includegraphics[width=70mm]{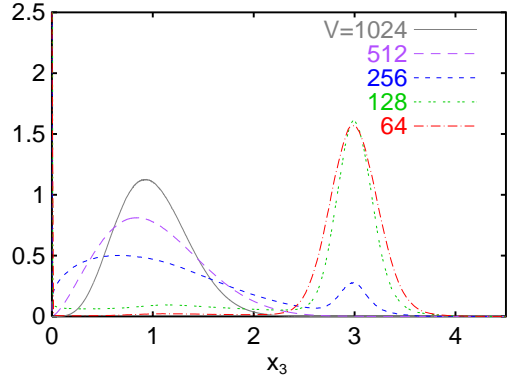}
\end{center}
\caption{Probability distribution of $x_{3}$ in
our model with $s_{1}=s_{2}=1.99$, $s_{3}=s_{4}=0.01$,
$D=1/128$, sampled over a time span of $5 \times 10^{6}$,
for different $V$ (see Case II for details).}
\label{fig:dist-x3}
\end{figure}

In the system investigated in our previous work,
the long-term average concentration of each chemical
does not differ from that in the continuum limit, since the system
over time switches between the
two states of broken symmetry.
It is important to determine if there are
systems for which the average concentrations of chemicals
are significantly altered by the DIT.
We will show that this is possible in a system
possessing some kind of asymmetry.  With the DIT without symmetry breaking,
the average concentrations of chemicals are drastically altered
by the change of $V$.
The peak position of the distribution is changed with a finite jump,
as the volume is decreased (see Fig. \ref{fig:dist-x3},
while see later sections for
the description of the model and simulation).
Borrowing the term of thermodynamics, 
the DIT reported previously is regarded as
a {\sl second order} transition involving symmetry breaking, while the DIT
reported here corresponds to the {\sl first order} transition without symmetry breaking.
This result is biologically significant as
providing a possible description of the alteration
of the concentrations of some
molecules within cells.  Note in a cell, the number of molecules of each chemical
species is not necessarily huge, and the discreteness effect is not always negligible.

To investigate this problem, we again use an autocatalytic reaction loop
of chemicals, but here we consider the case
in which $D_{i}$, $r_{i}$ or $s_{i}$ is
dependent on the chemical species $i$, where $r_{i}$ is the reaction constant
of the reaction $X_{i} + X_{i+1} \rightarrow 2X_{i+1}$.
To study the effects of discreteness,
we investigated the reaction model by using a stochastic particle simulation.
We assumed that the chemicals are well stirred in the container.
At each simulation step, two molecules in the container are randomly chosen.
Then we judge if the molecules react or not, by checking
if one of the two acts as a catalyst for the other as a substrate.
To carry out the simulation efficiently here, we adopted Gillespie's
direct method \cite{Gillespie1} (see Appendix) \cite{note0}.

Note that for there to be
a DIT to a 1-3 or 2-4 rich state,
it is necessary that
the time interval for inflow of molecules be longer than
the time scale of the reactions.
This time interval for $X_{i}$ inflow should be $\sim 1/D_{i}s_{i}V$.
In our previous study, in which we considered the case of identical parameter
values for all $i$, the
discreteness of the molecular concentration has the same effect
for all the molecule species,
and the transition occurs near $DsV = r$.

It is important to realize that
the relevance of the discreteness of the molecular concentration
depends on the reaction and flow rates of each molecule when the
parameter values are not identical.
For example, if $D_{1}s_{1} < D_{2}s_{2}$,
the inflow time interval for
$X_{1}$ molecules is longer than that for $X_{2}$ molecules,
so that the discreteness of the $X_{1}$ flow has a greater effect on
the behavior of the system. In general, $V$ may determine the
species that become extinct, and
the average concentration of each molecule can be greatly
changed by the discreteness effect.

Here, we consider the case in which
$s_{i}$ is species dependent, while $D$ and $r$ are identical for all
species, for the autocatalytic loop introduced above.
With this choice,
the discreteness effect of each chemical $X_{i}$ depends on $i$.
In this system, we find discreteness effects
that result in changes
of the average concentrations $\bar{x_{i}}$, with the temporal average
of the concentration $x_{i}=N_{i}/V$.
Although this result is obtained
with this simple example,
the mechanism we find would appear to be
quite general,
and hence there is reason to believe that the DIT we find exists in a
wide variety of real systems.

We first consider the effect of the discreteness of the inflow of chemicals
and how this depends on the relation between the reaction rate
and the inflow rate.
In our model,
the inflow interval of $X_{i}$ is $\sim \frac{1}{Ds_{i}V}$,
and the time scale of the reaction is $\sim \frac{x_{i}}{rx_{i}x_{i+1}}$.
When the former time scale is larger than the latter, the reaction from $i$ to
$i+1$ can proceed to completion before the inflow of species $i$ occurs.
Then $N_{i}$ becomes $0$. As long as $N_{i} = 0$, no reaction to produce
chemical $X_{i}$ occurs, and the average density may be
decreased radically from the continuum limit case.


\section*{Case I : inflow discreteness and reaction rate}

\begin{figure}
\begin{center}
\includegraphics[width=70mm]{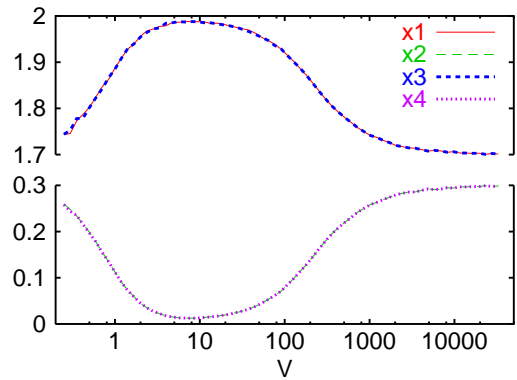}
\end{center}
\caption{The average concentration $\bar{x_{i}}$ in
Case I : $s_{1}=s_{3}=1.7$, $s_{2}=s_{4}=0.3$,
as a function of the volume $V$  (sampled over a time span
of $10^{6}$ for $V>1024$, $10^{7}$ for $32<V\le 1024$, and
$10^8$ for $V\le 32$,
also the same for Figs. \ref{fig:average11b} and \ref{fig:average11c}).
$D=1/128$.
}
\label{fig:average11a}
\end{figure}

As a simplest example to study this mechanism, we consider the case
with {\bf $s_{1}=s_{3} > s_{2}=s_{4}$}.
In this case, the rate equation in the continuum limit has a stable fixed
point $\forall i : x_{i} = s_{i}$.
When $V$ is large, each $x_{i}$ fluctuates around this fixed point.
The average concentration $\bar{x_{i}}$ is shown in
Fig. \ref{fig:average11a}.
With the decrease of $V$, the difference
between the pair $\bar{x_{1}}$ and $\bar{x_{3}}$
and the pair $\bar{x_{2}}$ and $\bar{x_{4}}$ is
amplified, and there is clear deviation from the continuum limit case.

\begin{figure}
\begin{center}
\includegraphics[width=70mm]{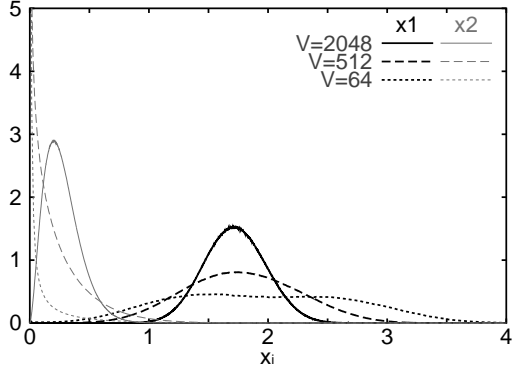}
\end{center}
\caption{Probability distribution of $x_{1}$ and $x_{2}$, sampled over
a time span of $5 \times 10^{6}$.
$s_{1}=s_{3}=1.7$, $s_{2}=s_{4}=0.3$ (Case I). $D=1/128$.
In the case $V=2048$, peaks around $x_{i}=s_{i}$, which correspond to
the fixed point at the continuum limit, is shown.
As $V$ decreases, the peaks get broader according to fluctuations,
and the tail of the distribution of $x_{2}$ reaches $0$.
Thus, there appears a peak around $x_{2}=0$.}
\label{fig:dist-a1}
\end{figure}

The mechanism responsible for this amplification
can be understood as follows.
As $V$ is decreased, we have found that the 1-3 rich state,
with extinction of
$N_{2}$ and $N_{4}$, appears when the increase of
$\bar{x_{1}}$ and $\bar{x_{3}}$ occurs.
To realize $N_{2}=N_{4}=0$,
it is necessary for the inflow interval of $X_{2}$ or $X_{4}$ to be
longer than the time scale of the reaction.
The inflow interval of $X_{i}$ molecules
is $\sim \frac{1}{Ds_{i}V}$,
while the time scale for the reaction is $\sim \frac{x_{i}}{rx_{i}x_{i+1}}$.
Since we set $r=1$ and $x_{i}=O(1)$, the 1-3 rich state appears for
$\frac{1}{Ds_{2}V},\frac{1}{Ds_{4}V} > \frac{1}{r}$, while
the 2-4 rich state appears for
$\frac{1}{Ds_{1}V},\frac{1}{Ds_{3}V} > \frac{1}{r}$.
Thus in the present case, the 1-3 rich state is first observed as
$V$ is decreased.
In the range of values of $V$ for which the relations
$\frac{1}{Ds_{1}V},\frac{1}{Ds_{3}V} < \frac{1}{r} < \frac{1}{Ds_{2}V},\frac{1}{Ds_{4}V}$ are satisfied,
the 1-3 rich state is realized often,
while the 2-4 rich state is not (see Fig. \ref{fig:dist-a1}).

Once the 1-3 rich state is realized, an $X_{2}$ molecule
and an $X_{4}$ molecule must enter the system
almost simultaneously for the system to break out of this state.
Thus the `rate of interruption' of the 1-3 rich state is roughly proportional
to $s_{2}s_{4}V^{2}$, the product of the rates of $X_{2}$ inflow
and $X_{4}$ inflow.
The expected residence time in the 1-3 rich state is the reciprocal
of the rate of interruption.
Thus the ratio of the expected residence times in the 1-3 rich and 2-4 rich
states is $\frac{s_{1}s_{3}}{s_{2}s_{4}}$.

From the above considerations, we expect that
for some $V\approx\frac{r}{Ds_{2}}$,
there appears a transition
to the 1-3 rich state, leading to a drastic increase of
the 1-3 concentration.
The validity of this conclusion has been confirmed by several
simulations, one of whose results is shown in
Fig. \ref{fig:average11a}.


\section*{Case I$'$ : imbalance of inflow discreteness}

The transition discussed above can
create a stronger effect on the concentrations.
As an example, consider the case
{\bf $s_{1}=s_{3} > s_{2} > s_{4}$}.

\begin{figure}
\begin{center}
\includegraphics[width=70mm]{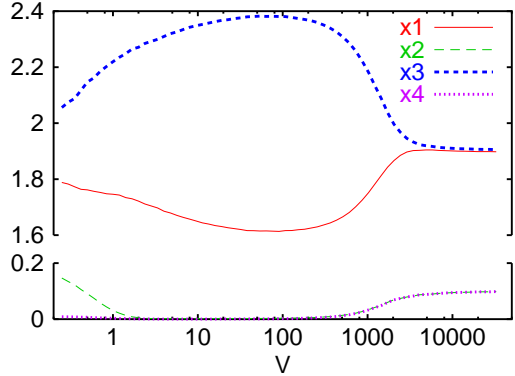}
\end{center}
\caption{The average concentration $\bar{x_{i}}$ in
Case I$'$ : $s_{1}=s_{3}=1.9$, $s_{2}=0.19$, $s_{4}=0.01$,
as a function of the volume $V$.
$D=1/128$.}
\label{fig:average11b}
\end{figure}

In this case, as in case I,
the 1-3 rich state is stable.
While in this state,
the system switches from a condition of
$N_{1} > N_{3}$ to one of $N_{1} < N_{3}$ due to
$X_{2}$ inflow and
from $N_{1} < N_{3}$ to $N_{1} > N_{3}$ due to $X_{4}$ inflow.
Since the latter event is less frequent
for $s_{2} > s_{4}$,
the condition $N_{1} < N_{3}$ is satisfied
for a greater amount of time in the 1-3 rich state.
Hence, it is expected
that $\bar{x_{1}} < \bar{x_{3}}$.
This is confirmed by the results displayed in
Fig. \ref{fig:average11b}.
This is in strong contrast with
the result in the continuum limit, where $\bar{x_{1}} \approx \bar{x_{3}}$
if $D \ll rs_{i}$ (i.e., the time scale of the reactions
is much shorter than that of the inflow).
The significant difference
between $\bar{x_{1}}$ and $\bar{x_{3}}$ found here appears
only when the 1-3 rich state is realized
through the effect of the discreteness of the
flow of $X_{4}$ molecules.
As shown in Fig. \ref{fig:average11b},
there is amplification of the difference between
$\bar{x_{1}}$ and $\bar{x_{3}}$ as $V$ decreases
that occurs simultaneously with
the transition to the 1-3 rich state.


\section*{Case II : inflow and outflow}

When $Ds_{i}V$ is small enough to insure the existence of
both 1-3 and 2-4 rich states,
the preference of states can depend on the
concentrations $s_{i}$.
The preferred state is selected through
another DIT caused by
outflow rather than
inflow of a particular chemical.

As an example, we consider the case {\bf $s_{2} \ge s_{1} > s_{3}=s_{4}$}.
Here again, the rate equation in the continuum limit has a stable fixed point.
If $D \ll rs_{i}$, then $x_{1},x_{3} \approx \frac{s_{1}+s_{3}}{2}$ and
$x_{2},x_{4} \approx \frac{s_{2}+s_{4}}{2}$ at the fixed point.

As discussed above, 1-3 and 2-4 rich states appear
for small $Ds_{3}V$.
In the 2-4 rich state,
it is likely for $N_{4}$ to decrease
as a result of the outflow of $X_{4}$ and
the reaction $X_{4} + X_{1} \rightarrow 2X_{1}$ facilitated
by the inflow of $X_{1}$.

If $s_{4}V < 1$, it may be the case that
all $X_{4}$ molecules flow out,
and $N_{4}$ becomes $0$.
The time required to realize $N_{4}=0$ from $N_{4}=n$ should be
$\sim \frac{1}{D} \log n$ when $s_{4}V \approx 0$ and $n \gg 1$.
However, if $s_{1}$ is large,
$X_{4}$ will be consumed by the reaction caused by $X_{1}$,
and for this reason, $N_{4}$ will decrease to $0$ more rapidly.
The time required to use up $X_{4}$ may also depend on $s_{1}$.
In this case, the 1-3 rich state is favoured by the mechanism
described below.

\begin{figure}
\begin{center}
\includegraphics[width=80mm]{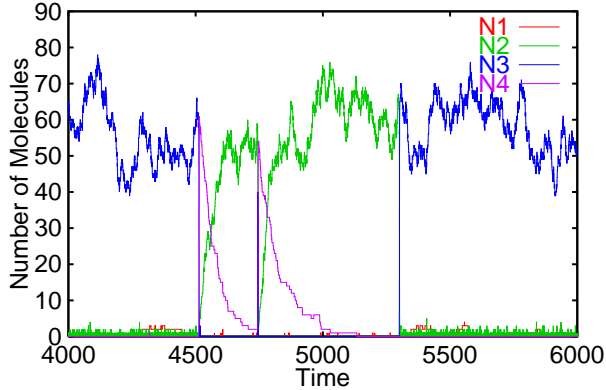}
\end{center}
\caption{Time series of the number of molecules $N_{i}$ for $V=16$.
Here $s_{1}=0.09$, $s_{2}=3.89$, $s_{3}=s_{4}=0.01$, and $D=1/64$.
There is a transition to the 2-4 rich state at $t=4511$.
In the 2-4 rich state, $X_{4}$ molecules flow out at the rate $D$, and
$N_{4}$ thereby decreases.
Due to the flow of $X_{3}$ molecules, switching from $N_{2} > N_{4}$ to
$N_{2} < N_{4}$ occurs, and $N_{4}$ increases again (as seen at $t=4743$).
Here, the interval over which the switching takes place is longer than
the interval of $X_{3}$ inflow, $1/DVs_{3}=400$,
and is indeed long enough for most $X_{4}$ molecules to diffuse out
before $X_{3}$ molecules can flow into the system.
Thus, here $N_{4}$ readily decreases to $0$ before the switch.
At $t=5300$, with $N_{4}=0$ an $X_{3}$ molecule flows into the container,
leading to a switch to the 1-3 rich state.}
\label{fig:ts}
\end{figure}

When $N_{4} > 0$,
$N_{4}$ can increase again as a result of $X_{3}$ inflow,
which leads to switch from the $N_{2} > N_{4}$ condition
to the $N_{2} < N_{4}$ condition.
The inflow interval for $X_{3}$ is $\sim \frac{1}{Ds_{3}V}$.
If this interval is much shorter than the time required
for $N_{4}$ to reach $0$, $N_{4}$ may increase again, causing the 2-4 rich
state to be preserved.
However, if the interval is longer,
$N_{4}$ may decrease to $0$, in which case, the 2-4 rich
state can be readily destroyed by the inflow of an $X_{3}$ molecule,
as shown in Fig. \ref{fig:ts}.

\begin{figure*}
\includegraphics[width=41mm]{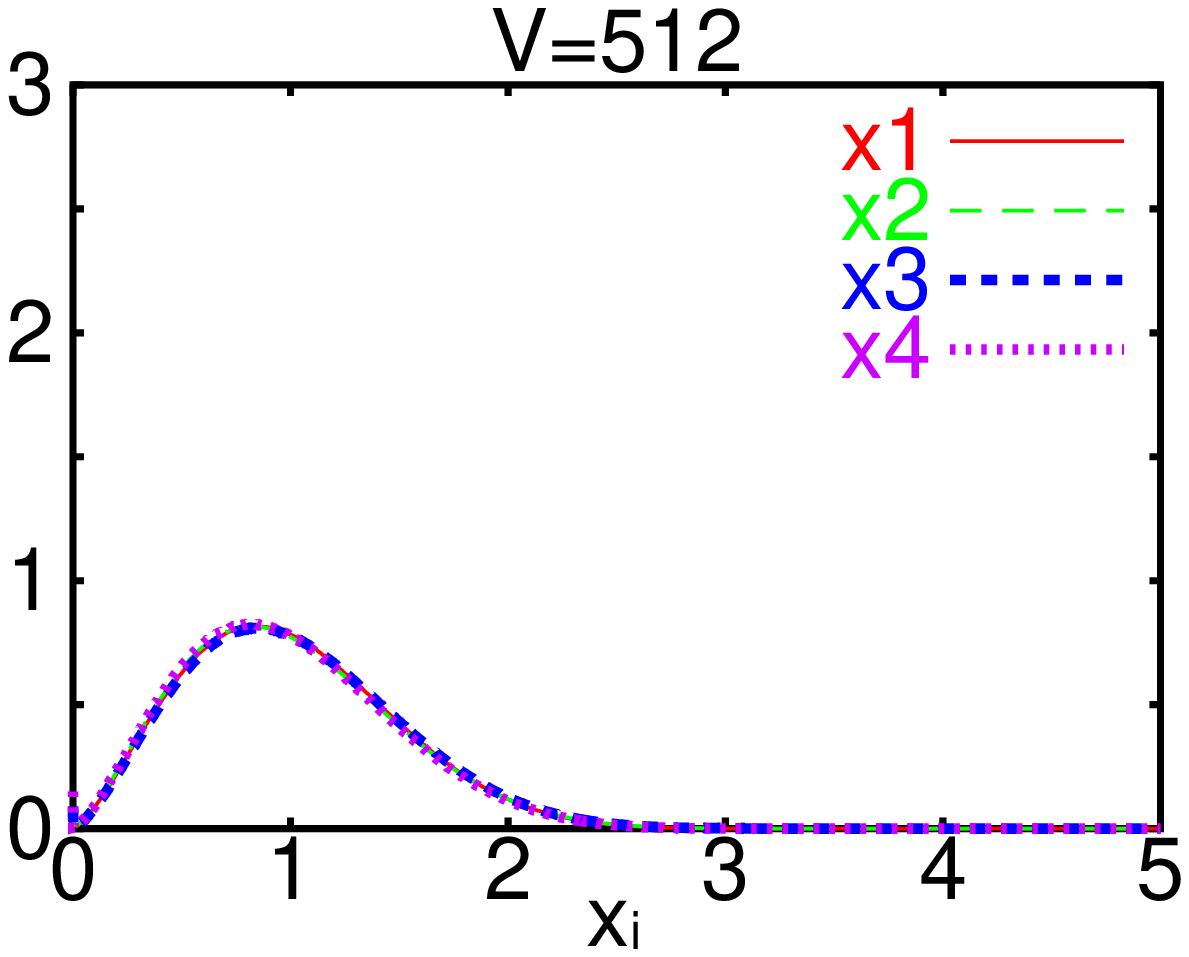}
\includegraphics[width=41mm]{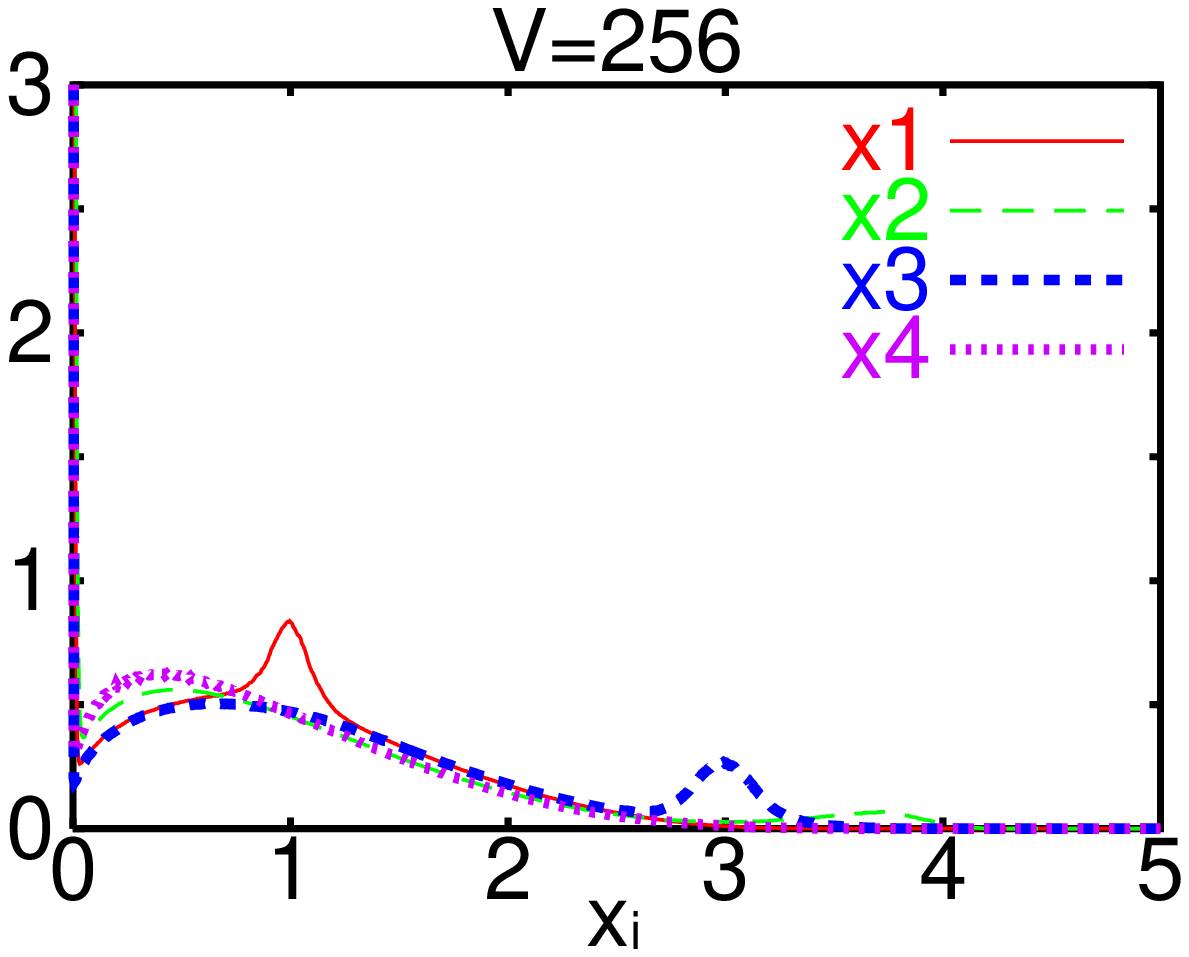}
\includegraphics[width=41mm]{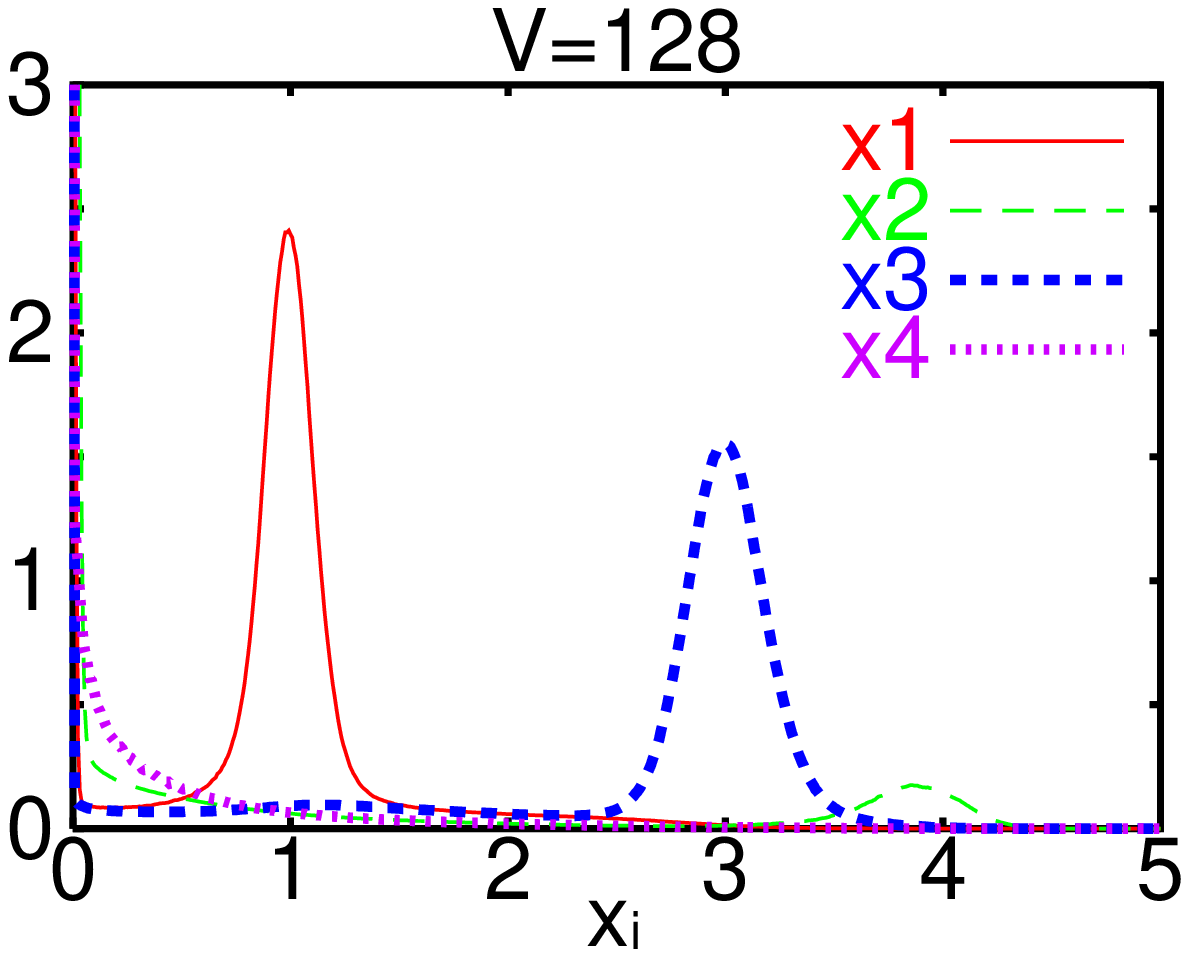}
\includegraphics[width=41mm]{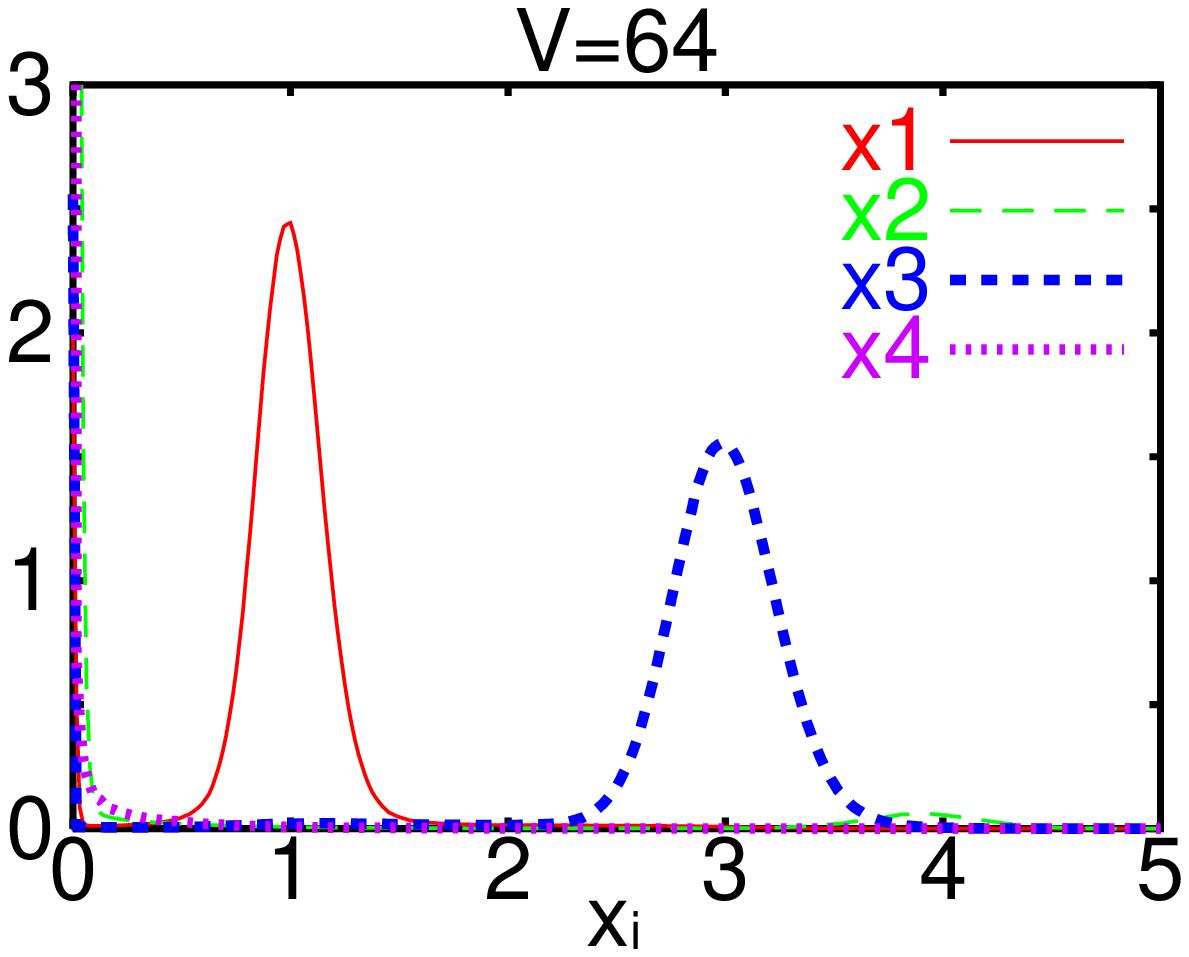}
\caption{Probability distribution of $x_{i}$ in
Case II : $s_{1}=s_{2}=1.99$, $s_{3}=s_{4}=0.01$,
sampled over
a time span of $5 \times 10^{6}$, for different $V$. $D=1/128$.
In the case $V=512$, all $x_{i}$ shows peaks around $x_{i}=1$,
which correspond to the fixed point at the continuum limit.
When $V$ is small, each $x_{i}$ shows a peak at a different concentration,
and the peak height
changes greatly with change of $V$.
For $V=256$, there appear peaks at around $x_{1}=1$, $x_{3}=3$, $x_{2}=4$, and
also around $x_{i}=0$ (for all $i$).
With further decrease of $V$, $N_{2}$ and $N_{4}$ reach $0$ more easily
than $N_{1}$ and $N_{3}$ do. Accordingly the peaks
for $x_{2}$ and $x_{4}$ around 0 grows as shown in  the case $V=128$.
The system tends to stay at the 1-3 rich state, and the peak around
$x_{2}=4$ gets smaller (as shown in the case $V=64$).}
\label{fig:dist-c1}
\end{figure*}

\begin{figure}
\begin{center}
\includegraphics[width=70mm]{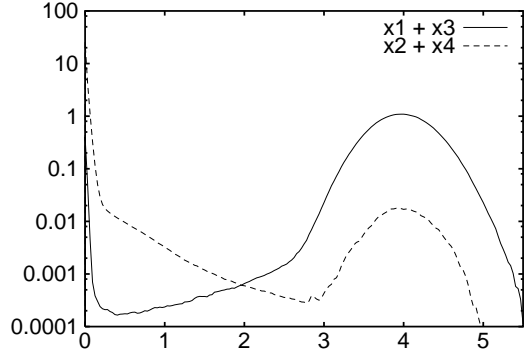}
\end{center}
\caption{Probability distribution of $(x_{1} + x_{3})$ and
$(x_{2} + x_{4})$ in
Case II : $s_{1}=s_{2}=1.99$, $s_{3}=s_{4}=0.01$,
sampled over a time span of $5 \times 10^{6}$.
$V=32$, $D=1/128$.
With such a small $V$, the 2-4 rich states are destabilized,
and the rate of the residence at the 1-3 rich states is
almost $10^{2}$ times larger than that at the 2-4 rich states.
The state allowed by the continuum limit, $x_{1}+x_{3}\approx x_{2}+x_{4}$ is
very rare.}
\label{fig:dist-c2}
\end{figure}

When the system is in the 1-3 rich state, on the other hand, switches
from a condition of $N_{1} > N_{3}$ to one of $N_{1} < N_{3}$ due to
$X_{2}$ inflow,
cause $N_{3}$ to remain large (as in case I$'$).
The system therefore tends to maintain
the condition $N_{1} < N_{3}$ (as long as $D/rs_{i}$ is not too large).
However, $N_{1}$ only rarely decreases to $0$, unlike $N_{4}$,
because $s_{1}$ is relatively large.
Thus the 1-3 rich state is more stable than the 2-4 rich
state, as shown in Figs. \ref{fig:dist-c1} and \ref{fig:dist-c2}.

\begin{figure}
\begin{center}
\includegraphics[width=70mm]{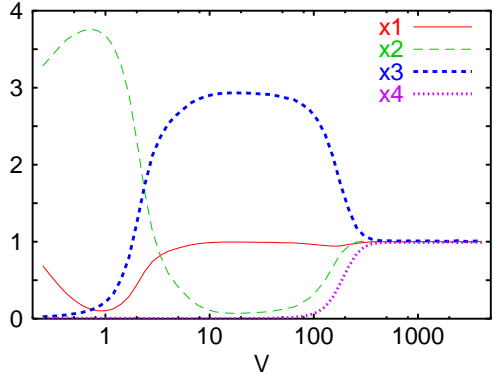}
\end{center}
\caption{The average concentration $\bar{x_{i}}$ in
Case II : $s_{1}=s_{2}=1.99$, $s_{3}=s_{4}=0.01$,
as a function of the volume $V$.
$D=1/128$.}
\label{fig:average11c}
\end{figure}

Hence, when $V$ is decreased sufficiently to satisfy
$s_{4}V < 1$, and the time interval $\frac{1}{Ds_{3}V}$
is sufficiently long to allow $N_{4}$ to decrease to $0$,
the 2-4 rich state loses stability, and the residence time
in the 1-3 rich state increases, due to the discreteness
of $X_{4}$.
As a consequence, $\bar{x_{3}}$ increases as $V$ decreases,
as shown in Fig. \ref{fig:average11c}.


\section*{Amplification by Discreteness}

Summarizing the findings discussed above,
differences among the `degrees of discreteness' of the chemicals
lead to novel DIT.
The average chemical concentrations are greatly altered by
this DIT.
Indeed, as the system size (the volume $V$) changes,
there is a sharp transition to a state
qualitatively different from that found in the continuum limit.
There are two key parameters
with regard to discreteness: One is $\frac{1}{Ds_{i}V}$ (investigated
in case I), the inflow time interval for $X_{i}$,
and the other is $s_{i}V$ (investigated in case II),
the number of species $X_{i}$ molecules in the system when it is at
equilibrium with the reservoir.

If the interval $\frac{1}{Ds_{i}V}$ is longer than the time
scale of the reaction, $\frac{x_{i}}{rx_{i}x_{i+1}}$,
the discreteness of the $X_{i}$ inflow is
relevant.
In such a situation, the $X_{i}$ molecules present in the system
may be completely consumed by the reaction
before any new $X_{i}$ molecules flow in, so that
$N_{i}$ may reach $0$.

Then, if the condition
$s_{i}V < 1$ is satisfied in addition to the above stated condition,
$N_{i}$ can become $0$ as a result of all $X_{i}$ molecules
flowing out of the system.
In this case, the relation between the time necessary to realize
a switch that increases $N_{i}$ and the time
necessary for $N_{i}$ to decay to $0$ is also important.

\begin{figure}
\begin{center}
\includegraphics[width=70mm]{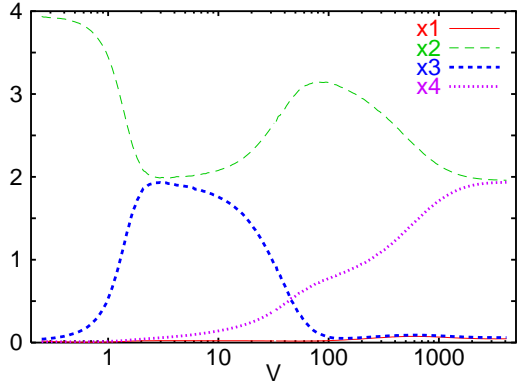}
\end{center}
\caption{The average concentrations $\bar{x_{i}}$,
sampled over a time span
of $5 \times 10^{8}$ for $V \le 32$ and $5 \times 10^{6}$ for $V > 32$,
plotted as functions of the volume $V$.
$s_{1}=0.09$, $s_{2}=3.89$, $s_{3}=s_{4}=0.01$, $D=1/64$.
For large $V$,
$\bar{x_{i}}$ is close to the fixed point value of the continuum limit.
As $V$ decreases, there first appears a 2-4 rich state,
but for smaller $V$, this 2-4 rich state becomes unstable,
and the residence time of the
1-3 rich state increases, leading to a sharp increase of $\bar{x_{3}}$.
For much smaller $V$ ($<0.5$),
only $X_{2}$ molecules exist for most of the time.}
\label{fig:average}
\end{figure}

With the above two conditions satisfied for each species $X_{i}$,
there appear several switches to different states
as $V$ is changed.
As an example, we considered the case in which
$s_{1}=0.09$, $s_{2}=3.89$, $s_{3}=s_{4}=0.01$, and $D=1/64$.
In this case, the average concentration $\bar{x_{i}}$ exhibits
three transitions as $V$ is decreased, as shown in
Fig. \ref{fig:average}.

\begin{figure}
\begin{center}
\includegraphics[width=70mm]{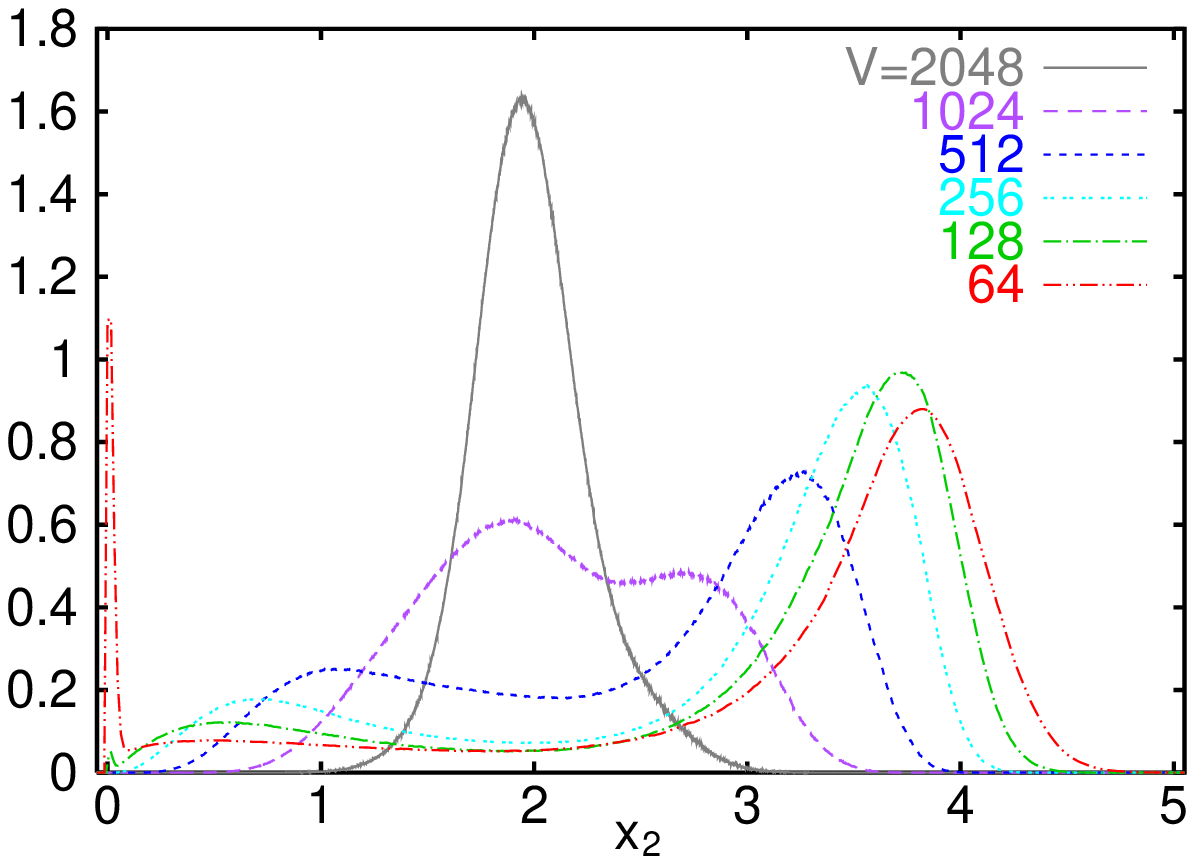}
\end{center}
\caption{Probability distribution of $x_{2}$, sampled over
a time span of $5 \times 10^{6}$.
$s_{1}=0.09$, $s_{2}=3.89$, $s_{3}=s_{4}=0.01$, and $D=1/64$.
When $V$ is large, there is a single peak around $x_{2}=2$,
which corresponds to the fixed point in the continuum limit.
Around $V=10^{3}$, double peaks
appear around $x_{2}=1$ and $x_{2}=3$, corresponding to the 2-4 rich state.
As $V$ decreases, these two peaks move apart,
and near $V=10^{2}$, the tail of the lower peak reaches $0$,
implying that $N_{2}$ and $N_{4}$ often decrease to $0$.
Thus here, the 2-4 rich state is unstable.
Then, as $V$ is decreased further,
the peak at $x_{2}=0$ rises sharply.}
\label{fig:dist-x2}
\end{figure}

First, in the continuum limit, $\bar{x_{1}}$ and $\bar{x_{3}}$ are very small,
as resulted from the fact
that $s_{1} + s_{3} \ll s_{2} + s_{4}$.
Around $V=10^{3}$, the discreteness of $X_{3}$ becomes significant,
and the 2-4 rich state appears.
Then the reactions $X_{2} + X_{3} \rightarrow 2X_{3}$ and
$X_{3} + X_{4} \rightarrow 2X_{4}$ take place only sporadically.
Contrastingly, the flow of $X_{1}$ molecules is fairly steady.
Thus, while the system is in the 2-4 rich state,
$N_{2} > N_{4}$ is satisfied for most of the time,
as shown in case I$'$.
Figure \ref{fig:dist-x2} displays the distribution of $x_{2}$.
Double peaks corresponding to the 2-4 rich state appear
in this situation.

In the 2-4 rich state with $N_{2} > N_{4}$,
$X_{2}$ molecules flowing into the system raise $N_{2}$ to the level
of $s_{2}V$, establishing equilibrium with the reservoir.
At the same time, $X_{4}$ molecules flow out,
and $N_{4}$ decreases to the level of $s_{4}V$, as seen in case II.

As seen in case II, The difference between $N_{2}$ and $N_{4}$ increases
with further decrease of $V$, since the switching rate decreases.
In Fig. \ref{fig:dist-x2},
the gap between the two peaks in the distribution of
$X_{2}$ is seen to become larger as $V$ decreases.
Around $V=10^{2}$, finally,
the imbalance between $N_{2}$ and $N_{4}$
destabilizes the 2-4 rich state.
For this reason, the 1-3 rich state becomes almost as stable as
(or more stable than) the 2-4 rich state,
in spite of the relation $s_{1} + s_{3} \ll s_{2} + s_{4}$.
The residence time in the 1-3 rich state increases sharply,
causing $\bar{x_{3}}$ to increase
(as shown in Figs. \ref{fig:ts} and \ref{fig:average}).
In fact, $\bar{x_{3}}$ increases to approximately $2$,
which is {\em more than 30 times larger}
than its value in the continuum limit \cite{note1}.

For very small $V$ (i.e. $V<2$), $N_{1}$ and $N_{3}$ decrease
to $0$ quite readily,
and thus the 1-3 rich state is also easily destroyed.
In this situation,
for most of the time only one chemical exists in the container.
Here, only $\bar{x_{2}}$ has a large value,
with all of the others near or at $N_{i}=0$.

In the manner described above,
non-trivial alteration of chemical concentrations
as a result of DIT was observed.
It has been found
that those molecule species whose numbers vanish are determined
not only by the flow rates
but also by the network and dynamics of the reactions.
For example, when $V$ is relatively large ($V\approx 10^{2}$),
$\bar{x_{3}}$ decreases as $s_{3}$ increases.


\section*{Discussion}

In conclusion,
we have reported a DIT that leads to
a strong effect on the average concentrations of the chemicals.
Although we have studied a simple case with only four chemicals here,
we have found that
this type of DIT appears in more complex reaction networks
of a more general nature.

In fact, we have randomly chosen a catalytic reaction network
consisting of few hundred species,
and studied the population dynamics of each chemical species with the
scheme of stochastic simulation adopted here.  For some reaction networks
we have observed the DIT as the volume is decreased.
In such cases, we have found that
the combination of the two mechanisms we studied here
leads to a variety of transitions
and alterations of molecular concentrations.
Although the example reported in the present paper is quite simple
and may look special, the mechanism found in the example gives a basis
for DIT in complex reaction network.

Generally speaking, DIT and its effects on molecular
concentrations are likely to be observed with chemical networks
containing autocatalytic reactions.
However, in some examples they are observed even without autocatalytic reactions.
When some part of reaction networks works as an autocatalytic sub-network
as a set, as seen in hypercycles \cite{Eigen}, the DIT of the present
mechanism is possible.

It is now experimentally feasible to construct a catalytic reaction
system in a micro-reactor,
and to design other types of systems with small numbers of molecules.
Also, there was great advance in techniques for detection
of small numbers (on the order of $1$ to $10^{2}$) of molecules
using fluorescence or other new methods such as thermal-lens
microscopy \cite{Kitamori}.
In such systems, experimental verification of DIT should be possible.
Also, we believe that the alterations of chemical concentrations
resulting from DIT that we found will have practical applications,
since quite high accumulation of dilute chemical species is possible
as we have shown here.

Since the number of molecules in a biological cell is often small,
the relevance of DIT to cell biology is obvious.
For example, in cell transduction, the number of signal molecules is
often less than 100, and even a single molecule
can switch the biochemical state of a cell \cite{McAdams}.
In our visual system,
a single photon in retina is amplified to a macroscopic level \cite{retina}.
Transmission of signals through neurons via synapses also
often involves a small number of molecules \cite{Bialek}.
Chemical reaction network consisting of several autocatalytic reaction
is widely seen in a cell, and such autocatalytic process  provides a candidate for
amplification of an effect of a single molecule.
Since the DIT we reported here is generally observed in autocatalytic
reaction networks, it is expected that it may be used in a biochemical reaction
network in a cell.
Indeed, according to our results,
the non-trivial accumulation of dilute molecules
and switching among several distinct states
with different chemical compositions may be realizable
by, for example, the control of flow by receptor.
Additionally, in some preliminary simulations with large reaction networks,
some sub-networks can be effectively activated or inactivated by DIT.
In such cases, transitions between several states characterized
by active sub-networks can be observed, which will be relevant to
switching between cellular states by a few signal molecules.

Switching the expression of genes on and off is a focus of interest
in bioinformatics.  This digital behavior is also
connected with the concentration of proteins present.
As is pointed out \cite{McAdams}, genetic regulation is
under stochasticity coming from smallness in the number of
associated molecules.
As we have seen in our model,
one chemical species can exhibit both an on/off switch and
continuous regulation of other chemicals,
even if the number of molecules of this species is small.
We believe that the switching of chemical states
facilitated by our DIT plays a role in
the regulation of genetic and metabolic processes in cells.

Throughout the paper we have adopted stochastic particle simulations.
Of course master equation approach is also equivalently possible, which
is especially useful if some analytic tools for it are developed.
For example, use of Fokker-Planck equations derived
in the limit of large volume (molecule numbers), is a powerful
tool \cite{Kampen}.  Since our DIT occurs when the volume
(the number of molecules) is quite small, such tools are so far
not available.  In future it will also be important to develop
some analytic tools for a system where the discreteness in the number is essential.

\subsection*{Acknowledgements}

This research was supported by
Grants-in-Aid for Scientific
Research from the Ministry of Education, Culture, Sports,
Science and Technology of Japan (11CE2006).


\section*{Appendix : Details of the Simulation}

\subsection*{Details of the Model}

We assumed that the chemicals are well stirred in the container and
the molecules have no volume.
Thus the rate of the reaction $X_{i} + X_{i+1} \rightarrow 2X_{i+1}$ is given by
$R_{i} \equiv r_{i}x_{i}x_{i+1}$ [concentration / time],
where $r_{i}$ is the reaction constant and $x_{i}$ is the concentration
of the chemical $X_{i}$.
By rewriting it with the use of $N_{i}$, the number of $X_{i}$ molecules, the rate of
the reaction is given by
$R_{i}V \equiv \frac{r_{i}N_{i}N_{i+1}}{V}$ [reactions / time].
In the same way, the rate of the $X_{i}$ inflow is given by
$D_{i}s_{i}V$ [molecules / time],
corresponding to $D_{i}s_{i}$ [concentration / time], whereas
that of the $X_{i}$ outflow is given by $D_{i}N_{i}$ [molecules / time],
corresponding to $D_{i}x_{i}$ [concentration / time].

This stochastic model approaches
the rate equation
\[
\frac{dx_{i}}{dt}=r_{i-1}x_{i-1}x_{i}-r_{i}x_{i}x_{i+1}+D_{i}(s_{i}-x_{i})
\label{eqn:1}
\]
when one takes a continuum limit, given by
$V \rightarrow \infty$.

We also assumed that the area of the surface of the container
is proportional to the volume $V$,
and thus the rate of the $X_{i}$ flow is proportional to $D_{i}V$,
to have this well-defined continuum limit for $V \rightarrow \infty$.
One might assume that the area of the surface should be $V^{2/3}$,
and the rate of the $X_{i}$ flow should be proportional
to $D_{i}V^{2/3}$.  This change of setting alters just the parameter values.
By suitably adjusting parameters $D_{i}$ and/or $s_{i}$,
the same transitions to the switching states and
the alteration of average concentrations are observed, even
with such settings.

With the rates of the reactions and the flows above,
we carried out the stochastic simulation.  In principle,
one can carry out the simulation, by randomly selecting two molecules, and
transforming one of them to other molecule, according to
the reaction rule, with the probability proportional to the
rate of reaction, when these molecules react.
Here, as an efficient simulation method, we adopt
Gillespie's direct method, instead.

\subsection*{Gillespie's Direct Method}

In our system, the state of the system is determined by $N_{i}$,
the number of molecules, and is changed only when one reaction or
one molecular flow occurs.
Thus the rate of the reactions and the flows do not change until
the next event (one reaction or one molecular flow) occurs, so that
the lapse time to the next event decays exponentially.

Gillespie's direct method \cite{Gillespie1} stands on this fact.
First, we determine the lapse time to the next event
by exponentially-distributed random numbers, and set the time forward.
Next, we determine which event occurs, with the proportion to the rate
of the event.
We change the state according to the event, and re-calculate
the rate of the reactions and the flows.
These steps are executed repeatedly, until the specified time elapses.

In some cases, especially with complicated reaction networks,
there are more efficient methods (see refs. \cite{Gillespie2,Gillespie3}).
Here, for simplicity, we adopted the Gillespie's direct method.
Our result discussed above does not depend on which method to use.





\begin{thebibliography}{99}
\bibitem{note0}
The choice of this method is only for efficiency.
The numerical result by the method agrees with that of the direct simulation,
as long as the time resolution and the resolution of the random number generator
is sufficient.
\bibitem{note1}
The amplification rate, i.e. the ratio of the maximum $\bar{x_{3}}(V)$,
to $\bar{x_{3}}$ at the continuum limit,
depends on the parameters $D$, $r$, and $s_{i}$.
At the continuum limit,
$\bar{x_{1}}, \bar{x_{3}} \approx (s_{1}+s_{3}) / 2$ and
$\bar{x_{2}}, \bar{x_{4}} \approx (s_{2}+s_{4}) / 2$,
if $D \ll rs_{i}$.
When $V$ is moderately small and the 1-3 rich states are more stable than
the 2-4 rich states,
$\bar{x_{1}} + \bar{x_{3}} > \bar{x_{2}} + \bar{x_{4}}$ holds,
as described already.
In this case, $\bar{x_{3}} \gg \bar{x_{1}}$ because $s_{2} \gg s_{4}$.
Thus the maximum value of $\bar{x_{3}}$ is around (or more than) half of
the total concentration $\sum{x_{i}}$ ($\approx \sum{s_{i}}$, if $D \ll rs_{i}$).
Thus, roughly speaking, the amplification rate of $\bar{x_{3}}$ is around
$(\sum{s_{i}}) / (s_{1}+s_{3})$.
For $s_{1}=0.09$, $s_{2}=3.89$, $s_{3}=s_{4}=0.01$,
the value is $40$, as is consistent with that shown in Fig. \ref{fig:average}.
\end{thebibliography}
\end{document}